\documentclass[letterpaper]{article}
\usepackage{emulateapj}
\usepackage{onecolfloat}
\usepackage{apjfonts}

\newcommand{\be}{\begin{equation}}
\newcommand{\ee}{\end{equation}}
\def\chan{{\sl Chandra}}
\def\asca{{\sl ASCA}}
\def\pdot{\dot{P}}

\def\ns{1E~1207.4--5209}

\begin{document}
\submitted{Accepted by ApJ Letters on March 12, 2002}
\twocolumn[
%
\lefthead{Pavlov et al.}
\righthead{Pulsar in the PKS 1209--51/52}
\title{1E 1207.4--5209: The puzzling pulsar
at the center of the PKS 1209--51/52 supernova remnant}
\author{
G.~G.~Pavlov\footnote{
The Pennsylvania State University, 525 Davey Lab,
University Park, PA 16802, USA; pavlov@astro.psu.edu},
V.~E.~Zavlin\footnote{
Max-Planck-Institut f\"ur Extraterrestrische Physik, D-85740
Garching, Germany; zavlin@xray.mpe.mpg.de}, 
D.~Sanwal$^1$, and J.~Tr\"umper$^2$}
\begin{abstract}
Second 
{\sl Chandra} observation of 1E 1207.4--5209, 
the central source of the  supernova remnant PKS
1209--51/52, allowed us to confirm the previously
detected period of 424 ms and,
assuming a uniform spin-down, estimate the period derivative,
$\dot P \sim (0.7$--$3)\times 10^{-14}$ s s$^{-1}$.
The corresponding characteristic age of the pulsar, 
$P/2\pdot \sim 200$--900 kyr, is much larger than the estimated
age of the SNR, $\sim 7$ kyr.  The values of the spin-down luminosity,
$\dot{E}\sim (0.4$--$1.6)\times 10^{34}$ erg s$^{-1}$, and
conventional magnetic field,
$B\sim (2$--$4)\times 10^{12}$ G, 
are typical for a  middle-aged radio pulsar, although no
manifestations of pulsar activity have been observed.
If \ns\ is indeed the neutron star formed in the same supernova
explosion that created PKS 1209--51/52, such a discrepancy in ages
could be explained either by a long initial period,  close to its
current value, or, less likely, by a very large braking index of
the pulsar. Alternatively, the pulsar could be a foreground object
unrelated to the supernova remnant, but the probability
of such a coincidence is very low.
\end{abstract}
\keywords{pulsars: individual (\ns) --- 
stars: neutron --- supernovae: individual (PKS 1209--51/52) 
--- X-rays: stars}

] 

\section{Introduction}
Radio-quiet compact central objects of supernova remnants
(SNRs) have emerged recently as a separate class of X-ray sources 
(see, e.g., Pavlov et al.\ 2002, for a review).  The nature of at
least some of these sources, presumably neutron stars (NSs),
remains enigmatic. They are characterized by soft, apparently thermal,
X-ray spectra and a lack of observed pulsar activity and optical
counterparts.  One of the most important properties, which potentially
allows one to elucidate the nature of these sources, is the periodicity
of their radiation.  Our previous observation of \ns, the 
central object of
PKS 1209--51/52 (=G296.5+10.0), with the Advanced CCD Imaging
Spectrometer (ACIS) on board of the {\sl Chandra} X-ray observatory
allowed us to detect a period $P=0.424129$ s (Zavlin et al.\ 2000;
Paper I hereafter), which proved that the source is indeed a 
neutron star.  
This source was discovered with the {\sl Einstein} observatory
(Helfand \& Becker 1984), $6'$ off the center of the $81'$
diameter SNR. From the analysis of radio and optical observations
of this SNR, Roger et al.\ (1988) estimated the SNR age to be 7 kyr,
with an uncertainty of a factor of 3. The distance to the SNR is
$d=2.1^{+1.8}_{-0.8}$ kpc (Giacani et al.\ 2000). The X-ray spectrum
of the central source can be described by a thermal model, with a
blackbody temperature of 3--4 MK and a radius of 1--2 km, at $d=2$
kpc (Mereghetti, Bignami, \& Caraveo 1996; Vasisht et al.\ 1997;
Zavlin, Pavlov, \& Tr\"umper 1998). Fitting the spectrum with a
hydrogen atmosphere model, Zavlin et al.\ (1998) obtained
an effective temperature of 1.4--1.9 MK and a radius of about 10--12 km.
Mereghetti et al.\ (1996) 
reported 
upper limits on the source flux
in radio ($< 0.1$ mJy at 4.8 GHz), optical ($V>23.5$),
and $\gamma$-rays ($<1\times 10^{-7}$ photons cm$^{-2}$ s$^{-1}$
for $E>100$ MeV). 

In this Letter we present the results of second
{\sl Chandra} ACIS observation, which allowed us 
to estimate the period derivative and other parameters
of the pulsar.
\section{Observation and data reduction}
\ns\ was observed with \chan\ on 2002 January 5--6  with the
spectroscopic array ACIS-S in the Continuous Clocking (CC) mode. 
This mode provides the highest time resolution of 2.85~ms available
with ACIS 
at the expense of one dimension of image.
The source was imaged on the back-illuminated chip S3, at a focal
plane temperature of --120~C.  The total duration of the observation
was 31.6~ks.  Time history of detected events does not reveal 
substantial background flares, so we do not exclude any time
intervals from the analysis.

The 1D image of \ns\ in ``sky pixels'' 
is consistent with
the ACIS Point Spread Function,  and it is similar to the image obtained
in the previous observation, taken at almost the same roll angle of
$65.2^\circ$ (vs.\ $65.3^\circ$ in the second observation). To obtain
the source count rate, we extracted 20,282  source-plus-background
counts from a 1D segment of 8 pixel length centered at the source
position (10 pixels contain 20,588 counts). The background  was taken
from similar segments adjacent to the 1D source aperture. Subtracting
the background, we find a source countrate of $0.64\pm 0.01$ s$^{-1}$,
versus $0.76\pm0.01$~s$^{-1}$ in the previous ACIS observation. 
The reduction of the count rate can be  attributed to a 
reduction of ACIS sensitivity at low energies.
\section{Timing analysis}
The event  times in the event file of a CC observation
are the times when the event is read out at the chip node.
To restore the actual event arrival times,
we subtracted the times of charge transfer 
between the detection point and the readout node
from the readout times
and corrected the times for the satellite wobbling (dither)
and the Science Instrument Module  motion 
using the approach described in Paper I.
The corrected times
were transformed 
to the solar system barycenter
using the {\tt axBary} tool of the CIAO package (v.\,2.2).

\begin{figure}[t]
\plotone{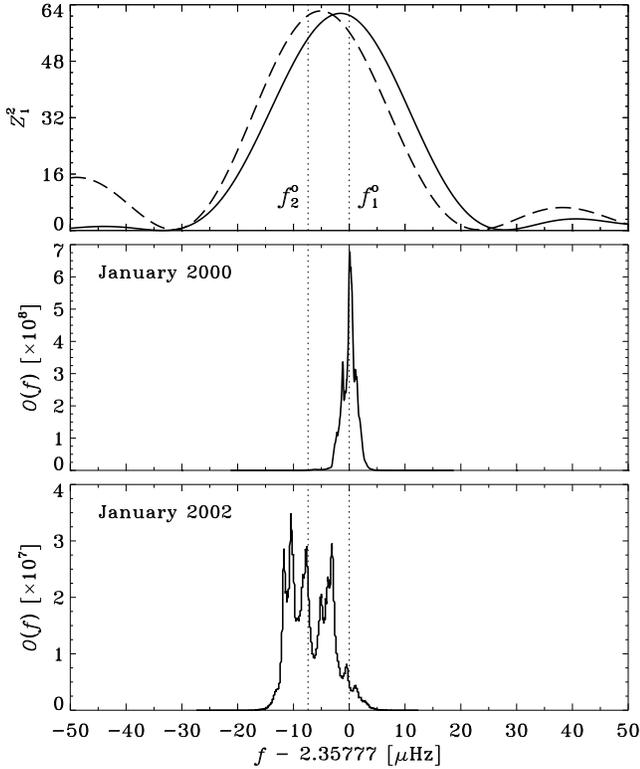}
\caption{Power spectra (upper panel) and frequency dependences of
odds ratio (middle and bottom panels) around the pulsation frequency
for the first and second observations (solid and dashed curves in the
upper panel, respectively). Dotted lines show the median
frequencies
$f^{\rm o}_{1,2}$ found in the two observations.
}
\end{figure}

 From a 3-pixel segment centered at the source position,
we extracted 18,237 counts 
in the 0.3--4.0~keV energy range, where the background contribution
is negligible (about $0.7\%$).
We ran the $Z_m^2$ test ($m$ is the number of harmonics
included --- see Buccheri et al.\ 1983)
 in the vicinity of the pulsation frequency
determined in our previous observation.
For $m=1$, 2, 3, 4, and 5,
we found the $Z_m^2$ peak values of 62.2, 63.9, 73.0, 77.6, and 79.3
at frequencies lower than a reference frequency 2.35777 Hz 
by 5.2, 4.9, 2.4, 2.8, and 3.3 $\mu$Hz, respectively.
The probability to
find such high peaks in a noise spectrum
is extremely low,
$\sim 10^{-14}$ in a frequency range $\la 10$ $\mu$Hz,
which confirms our previous period detection.
For a uniform comparison of the peak frequencies in first
and second observations, we recalculated the $Z_m^2$ peaks
for the first observation using exactly the same
extraction segments, energy range, and time corrections
as in the second one and obtained
$Z_m^2 = 61.6$, 61.8, 62.3, 62.7, and 65.6
(notice the smaller contributions from higher harmonics in first observation
compared to second one, in which third and fourth harmonics are
quite significant). The peak frequencies in the second observation are
lower than in the first one:
$f_1-f_2=4.2$, 2.9, 1.1, 1.9, and 2.2 $\mu$Hz,
for $m=1$, 2, 3, 4, and 5, respectively. 
This result shows that the pulsar has slowed down in two years, 
although the scatter of the frequency shift is rather high.

\begin{figure}[t]
\plotone{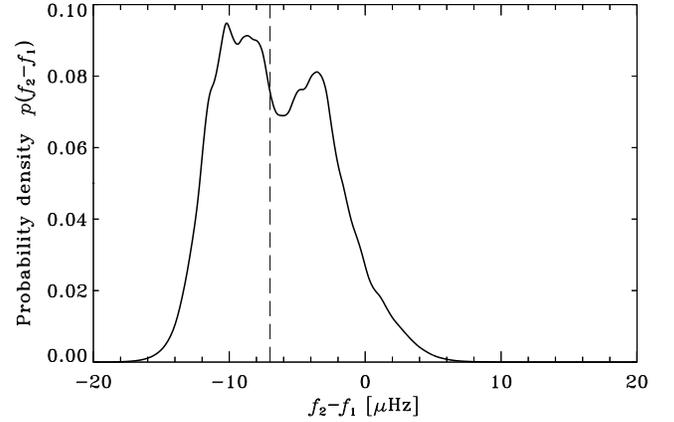}
\caption{Probability distribution for the frequency shift
(in $\mu{\rm Hz}^{-1}$). The vertical dash line shows the median
shift.
}
\end{figure}

To evaluate the most plausible frequencies and the frequency shift,
we employed the method of Gregory \& Loredo (1996), based on 
the Bayesian formalism. 
This method uses the phase-averaged epoch-folding algorithm to
calculate the frequency-dependent odds ratio $O(f)$.
This ratio specifies
how the data favor a periodic model of a given frequency $f$ over
the unpulsed model, and it allows one to find
 the corresponding probability distribution 
$p(f)\propto O(f)\, f^{-1}$ (see Gregory \& Loredo 1996,
Paper I, for details).  
The profile 
$O(f)$ calculated for the second observation appears to be
much broader than for the first one (FWHM$\sim 10$ $\mu$Hz
vs.\ $\sim 3$ $\mu$Hz), 
with several narrow peaks of
comparable heights (see Fig.\ 1). 
Exploring the $O(f)$ dependence in various energy ranges,
which show different relative contributions of the harmonics
of the pulsation frequency,
we found that the separate peaks are associated with different
harmonics (e.g., the highest peak in the lower panel of Fig.\ 1
is associated with $m=1$), in agreement with the different
$Z_m^2$ peak frequencies mentioned above.
The median frequency of the probability distribution in second 
observation, for the 0.3--4.0 keV band, is
$f_2^{\rm o}=2.3577625$ Hz; it differs from the mean frequency by
$\la 0.1$ $\mu$Hz.
The uncertainties of the frequency are $(-3.2,+4.5)$,
$(-4.2,+8.0)$, and $(-4.5,+10.1)$ $\mu$Hz at 68\%, 90\%, and 95\%
confidence level, respectively; the standard deviation is
$\sigma_{f2}=3.7$ $\mu$Hz.
We repeated a similar calculation for 
the data from the
first observation
(using the same extraction segment and energy range as for the
second one) and obtained $f_1^{\rm o}=2.3577699$ Hz,
with uncertainties of $(-1.4,+1.2)$, $(-2.5,+1.8)$, and
$(-2.9,+2.1)$ at 68\%, 90\%, and 95\%
confidence level, respectively;
$\sigma_{f1}=1.3$ $\mu$Hz.

We calculated the probability density distribution for the frequency difference,
$\Delta  = f_2-f_1$, 
as $p(\Delta) = \int p_2(f)\,\, p_1(f-\Delta)\,\, {\rm d}f$ (see Fig.\ 2),
where $p_1(f_1)$ and $p_2(f_2)$ are the probability densities
obtained for the two observations.
The median of this distribution, $\Delta^{\rm o} = -7.0$ $\mu$Hz,
gives a shift somewhat smaller than the mean shift, 
$\int \Delta\, p(\Delta)\, {\rm d}\Delta \simeq f_2^{\rm o}-f_1^{\rm o}
=-7.4$ $\mu$Hz, but the difference is well within the uncertainties
of $\Delta$: (--3.8,+4.5), (--5.4,+7.1), and (--5.9,+8.3) $\mu$Hz at 68\%,
90\%, and 95\% confidence level, respectively; $\sigma_\Delta = 3.9$ $\mu$Hz.
Because of the large widths of the $p_2(f_2)$ and,
consequently, $p(\Delta)$ distributions,
there is a non-negligible (albeit small) 5.6\% probability that $f_2>f_1$.

Adopting $\Delta^{\rm o} = - 7.0^{+4.5}_{-3.8}$ $\mu$Hz for the
frequency shift and assuming
a steady slowdown (lack of strong timing noise, glitches, etc)
during two years between the observations,
 we obtain the following estimates for the time
derivatives
\be
\dot{f}=-1.1^{+0.7}_{-0.6}\times 10^{-13}~{\rm Hz}~{\rm s}^{-1},\quad
\dot{P}=2.0^{+1.1}_{-1.3}\times 10^{-14}~{\rm s~s}^{-1}~.
\ee
The light curve extracted at $f=f_2^{\rm o}$ in the energy range
0.3--4.0 keV (Fig.\ 3, right panel)
reveals one broad pulse per period with 
a source pulsed fraction of
$f_{\rm p}=8\pm 2\%$, similar to that obtained in first observation.
 Figure 3 indicates that in both observations
there is a considerable shift of the pulse phase,
up to 0.4--0.6, between the energy bands 0.3--1.0 and 1.0--1.7 keV,
perhaps accompanied by a change of the pulse shape.
(Similar behavior has been observed in a number of middle-aged pulsars
--- e.g., \"Ogelman 1995.)
There is also some evidence that the pulse shapes are different in the
two observations --- e.g., in the 0.3--4.0 keV band
the pulse is closer to a sinusoid in first
observation while its shape becomes more
``triangular'' in second observation. 

\begin{figure}
\plotone{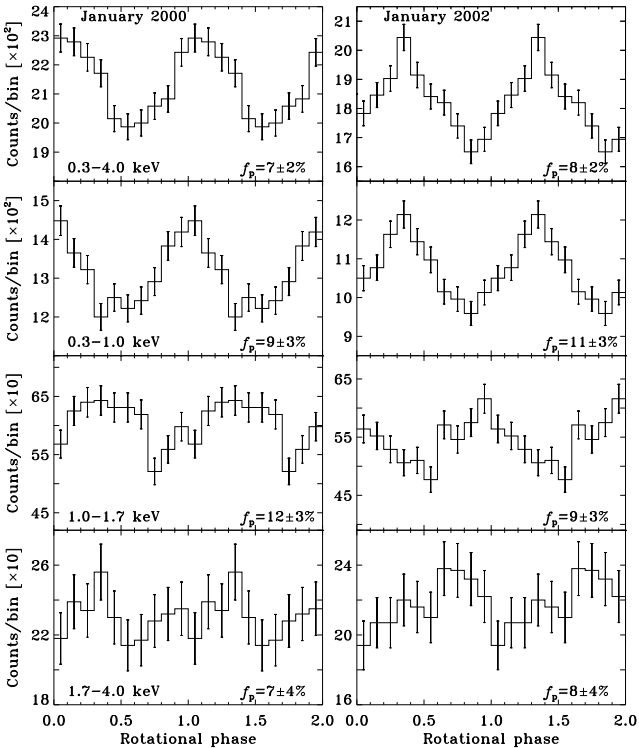}
\caption{Light curves extracted in different energy ranges for the
first and second observations (periods 0.4241296~s and 0.42413093~s,
respectively).}
\end{figure}

\section{Discussion}
The detection of approximately the same period in the second
observation confirms our previous detection and  proves unambiguously
that \ns\  is a 
neutron star. 
However, the time derivative of the
period appears to be surprisingly small ---  the corresponding
characteristic age of the pulsar,
$\tau_c = P/(2\dot{P})=340^{+600}_{-120}$ kyr,
exceeds the estimated SNR age, 3--20 kyr, by a large factor, and even
the most conservative upper limit, 
$\pdot < 5\times 10^{-14}$~s~s$^{-1}$,
yields an incredibly old age, 
$\tau_c > 130$ kyr.  Such values of $P$ and $\pdot$, as well as the
corresponding values of rotation energy loss rate,
$\dot{E}=4\pi^2 I \pdot P^{-3}=
1.0^{+0.6}_{-0.7}\times 10^{34} I_{45}$ erg s$^{-1}$,
and 
conventional ``magnetic field'',
$B=(3Ic^3P\dot{P}/8\pi^2R^6)^{1/2} = 
2.9^{+0.8}_{-1.1}\times 10^{12} I_{45}^{1/2}R_6^{-3}$ G,
 are typical for radio pulsars.

This result could be easily explained if one assumes
that \ns\ is merely a middle-aged field pulsar, unrelated to the SNR.
If this pulsar were a background object at a distance well
beyond the SNR, it would be difficult to explain
the very high (thermal) X-ray luminosity 
--- e.g., $L_{\rm x} \approx 3\times 10^{34}$ erg s$^{-1}$
at $d=10$ kpc cannot be emitted from radio pulsar polar caps
because it exceeds $\dot{E}$, 
and it cannot be emitted from the surface
of a cooling NS because the observed temperature is too high
for such an age.
One, however, might assume that the pulsar is a foreground object,
at a distance of several hundred parsecs,  perhaps similar to
the radio-quiet Geminga ($\tau_c\sim 300$ kyr, $d\sim 0.2$ kpc)
or (radio-bright) PSR B1055--52 ($\tau_c\sim 500$ kyr, $d\sim 1$ kpc).
In this case, the  observed thermal-like pulsed X-ray
radiation could be interpreted as emitted from hot polar caps
with a size of a few hundred meters.
The lack of 
detected
radio and $\gamma$-ray radiation might be due to
unfavorable orientation of pulsar beams
(albeit $\gamma$-ray beams are expected to be rather broad). 
To explain the lack of softer thermal radiation from the entire
cooling NS surface (like that observed from Geminga and
PSR~B1055--52),  one could speculate
that \ns\ is older than the above-mentioned pulsars
(e.g., $\tau_c \sim 1$ Myr cannot be ruled out),
so that its surface temperature is lower, and its
thermal radiation is too soft to be observable in X-rays.
However, although the observed properties of \ns\
cannot exclude the hypothesis that it is a foreground pulsar,
the probability to find a middle-aged pulsar so close to the
line of sight towards the SNR center is very low. Therefore, 
we consider this interpretation unlikely.

If \ns\ is indeed associated with the SNR,
we can reconcile the pulsar's age with that of the SNR,
assuming that the braking index $n$ of the pulsar is large
and/or its initial period $P_0$ was close to its current value.
If the pulsar slowdown is described by the equation $\dot{f}=-K f^n$,
then its age, for a constant $K$,
is
\be
\tau = 
\frac{P}{(n-1)\dot{P}}\left[1-\left(\frac{P_0}{P}\right)^{n-1}\right].
\ee
To obtain $\tau\sim 20$ kyr (an upper limit on the age of
PKS 1209--51/52), one can assume, for instance, $n\sim 20$--90
at $P_0\ll P$,  or 
$P_0\simeq 395$--417 ms for $n=2.5$. We are not
aware of physical models of pulsar deceleration which would
give so large braking indices, although Johnston \& Galloway
(1999) argue that empirical braking indices of relatively
old pulsars can be as large as 20--50. On the other hand,
there is no physical limitations on the initial period ---
it can be very close to its current value if the constant $K$
in the slowdown equation is small. 

It should be noted that \ns\ is not the only pulsar that
shows a characteristic age much older than the age of the
SNR with which it is apparently associated. A similar
example is the 7-s Anomalous X-ray Pulsar (AXP)
1E 2259+586  associated with the 
SNR CTB 109 --- the SNR age, $\sim 3$--21 kyr
(Rho \& Petre 1997; Parmar et al.\ 1998),
is much younger than
$\tau_c\simeq 226$ kyr (Kaspi, Chakrabarty, \& Steinberger 1999).
Another example is the
65~ms rotation-powered X-ray pulsar AX~J1811.5--1926, whose period
and period derivative were measured with \asca\ 
(Torii et al.~1997, 1999), and association with the remnant
G11.2--0.3 of the historic supernova A.D.~386 was
strongly supported by \chan\ observations
(Kaspi et al.~2001a). The characteristic age of AX~J1811.5--1926,
about 24~kyr, is 15 times the true age. 

If the pulsar was born in the supernova explosion
that created PKS 1209--51/52, its X-ray luminosity,
$L_{\rm x}\approx 1\times 10^{33}$ erg s$^{-1}$ in the 0.5--6.0 keV band
(at $d=2$ kpc), is 
0.06--0.3 of the spin-down luminosity $\dot{E}$.
This means that \ns\ is different from AXPs, which show $L_{\rm x}>\dot{E}$.
On the other hand, the fraction  of the spin-down luminosity detected
in X-rays is too large to interpret the X-ray radiation  as a nonthermal
radiation 
or thermal polar-cap radiation emitted from a rotation-powered
pulsar
(in fact, the X-ray spectrum of \ns\ is inconsistent with
a nonthermal, power-law model
--- see Zavlin et al.~1998).
Therefore, the original interpretation of the spectrum as emitted
from the surface of a cooling NS (Matsui, Long, \& Tuohy 1988) remains
most plausible, and the size of the emitting region and the effective
temperature can be reconciled with those expected for a 10-kyr NS
assuming that the NS is covered with a hydrogen or helium atmosphere
(Zavlin et al.\ 1998).
The magnetic field $\sim 3\times 10^{12}$ G, inferred from the
estimated $\pdot$ assuming a centered magnetic dipole and
$R_{\rm NS}=10$ km,  may look too low to cause the
temperature non-uniformity  (Shibanov \& Yakovlev 1996)
required to explain pulsations detected from \ns. However, we can
speculate that the magnetic dipole is off-centered,
which  can give a much stronger local magnetic field at the
same value of the magnetic moment. 

While estimating the period derivative from  just two observations,
one should not forget that such an estimate implies a lack of
glitches or other timing irregularities. Young and middle-aged
radio pulsars show glitches with $\delta f$ up to $5\times 10^{-6}f$
(Lyne \& Graham-Smith 1998). Even higher timing irregularities have
been observed from some AXPs (e.g., $\delta f\sim 10^{-4}f$ in
1E~1048.1--5937 --- Kaspi et al.\ 2001b).  If, for instance, our
pulsar had a glitch  $\delta f \sim 10^{-5}f$  just before the
second observation, the pre-glitch frequency would be lower than
the measured one by $\sim 20$ $\mu$Hz, which would increase 
the frequency shift
and lowered the NS age by a factor of 3. 
Finally, we cannot rule out that \ns\ is in a binary system with a long
orbital period --- our X-ray observations, each spanning about 8
hours, suggest that the period should be
$P_{\rm orb}\ga 40$ hours, as no effect of the binary orbit is
clearly observed in the data, and the optical observations were not
deep enough to exclude the presence of a faint secondary companion. 
In this case, the X-ray observations give Doppler-shifted periods
--- e.g., a radial velocity of 10 km s$^{-1}$ corresponds to a shift
of 79 $\mu$Hz, much larger than the observed one.
Therefore, the above estimates
of pulsar parameters should be taken with caution
until more timing observations of the pulsar are carried out.

To conclude, we measured the shift of the pulsar period between two
{\sl Chandra} observations. The period and its apparent derivative
do not contradict to the hypothesis that it is a young pulsar
physically associated with the SNR PKS 1209--51/52.
 From the data available, we cannot completely rule out that 
that this is a foreground middle-aged
pulsar, but this 
alternative interpretation looks less plausible.  The inferred
properties of the pulsar warrant deep radio, optical, and $\gamma$-ray
observations of this object.
To understand the nature of the pulsar, its timing
behavior should be monitored in X-rays, which would allow one to
find out whether it shows strong timing irregularities
or it is in a binary system.  Another clue to the nature of \ns\ would
be provided by deep, high-resolution spectral observations of this
object in X-rays, which could detect and identify spectral lines and
evaluate the surface magnetic field of the 
neutron star.

\acknowledgements
We thank Glenn Allen and Allyn Tennant for the helpful advice
on the ACIS timing issues.
We are grateful to Marcus Teter and Fernando Camilo for useful discussions.
This work was partly supported
by SAO grant GO2-3088X and NASA grant NAG5-10865.
{}
%
%

\end{document}